\documentclass[english]{JHEP3}

\usepackage{amsmath,cite,babel}
\usepackage{graphicx,axodraw}
\usepackage{array}
\usepackage{xspace}

\newcommand{\mr}{\mathrm}
\newcommand{\gev}{\,GeV}

\newcommand{\mrgev}{\mr{\gev}}

\newcommand{\me}{\mathcal{M}}
\newcommand{\mee}[1]{\overline{|\me_{#1}|}^{\,2}}
\newcommand{\invPA}[2]{\left(#1^2-#2^2\right)^2+ #1^2\Gamma^2(#1)}

\newcommand{\ie}{{\it i.e.}\xspace}
\newcommand{\eg}{{\it e.g.}\xspace}
\newcommand{\fortran}{\textsf{FORTRAN}\xspace}
\newcommand{\HW}{\textsf{HERWIG}\xspace}
\newcommand{\HWPP}{\textsf{Herwig++}\xspace}
\newcommand{\sla}[1]{#1\!\!\!\!\!\!\not\,\,\,}

\title{Simulation of Finite Width Effects in Physics 
Beyond the Standard Model}

\author{M.A.~Gigg \\ IPPP, Department of Physics, Durham University\\
  Email: \email{m.a.gigg@durham.ac.uk}}
\author{P.~Richardson, \\ IPPP, Department of Physics, Durham University; and\\
  Physics Department, CERN \\
  Email: \email{peter.richardson@durham.ac.uk}}


\preprint{DCPT/08/68 \\ IPPP/08/34 \\ CERN/PH/TH-110 \\ MCnet/08/01}

\abstract {
  We present a method for the inclusion of finite width effects in the simulation of 
  Beyond Standard Model~(BSM) physics. In order to test the validity of the method 
  we compare our results with matrix elements for a range of 
  production and decay processes in the Standard Model, Minimal Supersymmetric Standard 
  Model~(MSSM) and Minimal Universal Extra Dimensions model~(MUED). This procedure has 
  been implemented in the \HWPP event generator and will be available in a 
  forthcoming release.
}

\begin{document}

\section{Introduction}
\label{sec:intro}
It has been believed for some time that there must be new physics
at the TeV scale. While most theories of physics
beyond the Standard Model differ in their internal workings they all possess 
a common feature, the appearance of a spectrum of heavy states 
that will be produced at a collider with sufficient energy. If there is to be
any hope of discovering these new particles one must have accurate simulations 
of the signals they will produce, and in particular, accurate simulations that
can be compared directly with experimental data. There are, essentially,
two methods of producing events distributed as they would be in an experiment. The
first is to use a matrix element generator~\cite{Maltoni:2002qb,Pukhov:1999gg,Pukhov:2004ca,Kilian:2001qz,Moretti:2001zz,Krauss:2001iv} to 
compute the full $2\to n$ matrix element for a specific final state of interest, 
including the decays of all of the fundamental particles. This can
then be interfaced to a general-purpose event generator to produce the 
particles observed in an experiment. The advantages are that the 
full matrix element calculation gives
contributions from non-resonant diagrams which may be important and the effects of
spin correlations are automatically included. Ideally, we would use an
$n$-body matrix element for all processes, however, there are a number of issues:
\begin{enumerate}
  \item the time to compute the full matrix element, even if only the 
    resonant diagrams are included, grows with the number of final-state 
    particles. This is particularly problematic for models such as supersymmetry
    which can contain long decay chains;
  \item many new physics models introduce a large number of 
    new states, which lead to many possible production and decay mechanisms. If the 
    full calculation is used in all cases the number of processes 
    required becomes prohibitively large;
  \item any new coloured states will emit QCD radiation in their
    production and decay which must be simulated using the parton shower 
    approximation.
\end{enumerate}

Given these issues, for the foreseeable future, we will need to use general purpose
event generators~\cite{Bahr:2008pv,Corcella:2000bw,Moretti:2002eu,Sjostrand:2006za,Sjostrand:2007gs,Gleisberg:2003xi} that treat the 
production and decay of heavy particles in a factorised approximation. If a generator
incorporates the model of interest it can simulate everything from the initial 
hard interaction up to the final colour-singlet particles that interact with 
a detector. The production step is typically a simple $2\to 2$ scattering followed
by a series of perturbative decays which result in stable particles that are
hadronized and decayed to give the observed colour-singlet states\footnote{This is an extremely simplified schematic of event generation. More detail can be found in, for
example, Ref.~\cite{Bahr:2008pv}.}. It is vital that these simulations are as 
accurate as possible by, for instance, including effects such as spin 
correlations~\cite{Knowles:1988vs,Collins:1987cp,Richardson:2001df}
throughout the event simulation.

Another area where the factorized approach can neglect important physics is in the 
treatment of off-shell effects, as by definition the method must use the narrow width
approximation to separate production and decay. In the simplest
approach all particles are on-shell throughout their production and 
decay~\cite{Gigg:2007cr}, which in many scenarios is a good enough approximation but
in other regimes, \eg close to thresholds and for resonances, such as the $Z^0$
boson, where the width can be measured, this approach is unsatisfactory. In order
to improve the physics of the simulation, off-shell effects must be included in the 
production and decay stages of the event generation. The aim of this paper
is to describe how this can be achieved and discuss its implementation in the 
\HWPP event generator~\cite{Bahr:2008pv}.

The next section describes the narrow width approximation in more detail, 
Sect.~\ref{sec:offshell} gives a description of the method by which we include
width effects, with specific examples for the production and decay stages.
Section~\ref{sec:examples} gives results for various scenarios in the 
Minimal Supersymmetric Standard Model~(MSSM) and the Minimal Universal Extra
Dimensions~(MUED) model, some conclusions are drawn in the final section. 
 
\section{Narrow Width Approximation}
\label{sec:nwa}
In general the evaluation of the matrix element for a process with a high 
multiplicity final state is complex due to the factorial growth in the number
of diagrams. The calculation can be simplified by using the narrow width
approximation, where if:
\begin{enumerate}
  \item the resonance has a small width $\Gamma$ compared with its pole mass $M$, 
    $\Gamma \ll M$;
  \item we are far from threshold, $\sqrt{s} - M \gg \Gamma$, where $\sqrt{s}$
    denotes the centre-of-mass energy;
  \item the propagator is separable;
  \item the mass of the parent is much greater than the mass of the decay products;
  \item there are no significant non-resonant contributions;
\end{enumerate}
the cross section integral can be separated into an on-shell production step
followed by a series of decays. The separation arises from integrating
out the propagators connecting each step giving a momentum independent factor
\begin{equation}
\label{eqn:nwaint}
  \int_{-\infty}^{\infty}\mr{d}q^2 \left|\frac{1}{(q^2-M^2) +iM\Gamma}\right|^2
  = \frac{\pi}{M\Gamma}.
\end{equation}
If the above assumptions are true one obtains an estimate of the
cross section with an error of $\mr{O}(\Gamma/M)$ using this approach.

In reality, especially when dealing with BSM physics, the approximation is 
commonly used when the above assumptions are not strictly satisfied. While there
have recently been studies of the validity of narrow width limit for some
SUSY scenarios~\cite{Berdine:2007uv,Kauer:2007zc,Kauer:2007nt}, nothing has
been studied in relation to other popular new physics models. In the next
section we describe how off-shell effects can be simulated with our main 
focus on BSM studies. In \HWPP the same approach is used for the simulation
of off-shell effects in hadron decays, allowing the same infrastructure to handle
all decays.

\section{Off-Shell Weight Factor}
\label{sec:offshell}
As discussed above, the narrow width approximation allows the propagator
connecting production and decay of successive decays to be integrated out. This 
essentially means that part of the phase-space integral is approximated
to a constant when the correct assumptions are satisfied. Therefore, 
to improve the accuracy of our simulation we wish to move away from the on-shell
approximation and include the effects from integrating over the 
connecting propagator. In the past this has been accomplished in a variety
of ways. For example, the \fortran \HW~\cite{Corcella:2000bw} program included:
\begin{enumerate}
\item the full three-body matrix element, with an off-shell $W^\pm$ boson, 
  for top decay;
\item smearing of fundamental particle masses using a Breit-Wigner distribution;
\item a more sophisticated Higgs boson lineshape~\cite{Seymour:1995qg}.
\end{enumerate}

To improve our simulation we include the weight factor~(see Sect.~\ref{app:wgtderiv} 
for a derivation)
\begin{equation}
  \label{eqn:ofswgt}
  \frac{1}{\pi}\int_{m^2_{\mr{min}}}^{m^2_{\mr{max}}}\mr{d}m^2 \frac{m\Gamma(m)}
       {(m^2-M^2)^2 + m^2\Gamma^2(m)},
\end{equation}
throughout the production and decay stages, where $\Gamma(m)$ is the running 
width of the particle, $M$ is the pole mass and $m_{min,max}$ are defined 
such that the maximum deviation from the pole mass is a constant times the 
on-shell width.
The weight includes a momentum dependence into the calculation of 
cross sections and decay widths thereby improving the approximation to the 
full matrix element. While for the latter case convoluting the weight with 
the partial width calculation for a particular decay mode is relatively 
simple, this is not the case for the production stage, so we will 
consider them separately.

\subsection{Off-shell Masses in Particle Production}
For production we need to convolute the weight factor described above with the
cross section integral. We achieve this by distributing the masses of the outgoing 
particles according to Eq.~(\ref{eqn:ofswgt}). This in itself is a 
trivial task but what we will show here, with an example from supersymmetry, is 
that gauge invariance can be violated if these masses are used na\"ively 
when calculating the matrix elements.

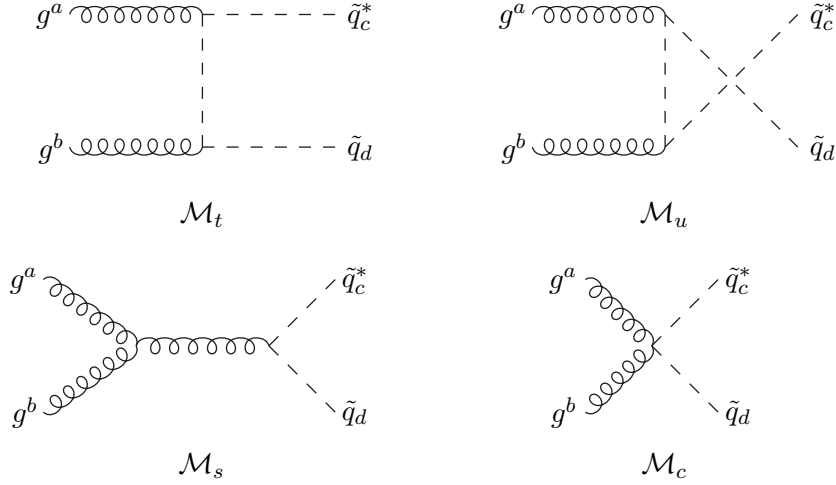
\begin{figure}
  \centering{
    \begin{picture}(300,200)
      \Gluon(5,195)(55, 195){3}{7} \Text(3,195)[r]{$g^a$}
      \DashLine(55,195)(105,195){5} \Text(110,195)[l]{$\tilde{q}^*_c$}
      \DashLine(55,195)(55,145){5}
      \Gluon(5,145)(55,145){-3}{7} \Text(3,145)[r]{$g^b$}
      \DashLine(55,145)(105,145){5} \Text(110,145)[l]{$\tilde{q}_d$}
      \Text(55,120)[c]{$\me_t$}
      \SetOffset(175,0)
      \Gluon(5,195)(55, 195){3}{7} \Text(3,195)[r]{$g^a$}
      \DashLine(55,145)(105,195){5} \Text(110,195)[l]{$\tilde{q}^*_c$}
      \DashLine(55,195)(55,145){5}
      \Gluon(5,145)(55,145){-3}{7} \Text(3,145)[r]{$g^b$}
      \DashLine(55,195)(105,145){5} \Text(110,145)[l]{$\tilde{q}_d$}
      \Text(55,120)[c]{$\me_u$}
      \SetOffset(-10,-10)
      \Gluon(5,105)(40,80){3}{6} \Text(3,105)[r]{$g^a$}
      \Gluon(5,55)(40,80){-3}{6} \Text(3,55)[r]{$g^b$}
      \Gluon(40,80)(90,80){3}{6}
      \DashLine(90,80)(115,105){5} \Text(118,105)[l]{$\tilde{q}^*_c$}
      \DashLine(90,80)(115,55){5} \Text(118,55)[l]{$\tilde{q}_d$}
      \Text(65,35)[c]{$\me_s$}
      \SetOffset(185,-10)
      \Gluon(15,105)(40,80){3}{5} \Text(12,105)[r]{$g^a$}
      \Gluon(15,55)(40,80){-3}{5} \Text(12,55)[r]{$g^b$}
      \DashLine(40,80)(65,105){5} \Text(68,105)[l]{$\tilde{q}^*_c$}
      \DashLine(40,80)(65,55){5} \Text(68,55)[l]{$\tilde{q}_d$}
      \Text(45,35)[c]{$\me_c$}
    \end{picture}
    \vspace{-8mm}
  }  
  \caption{Feynman diagrams for the process $gg\to \tilde{q}^*\tilde{q}$
    where the Roman indices give the colour representation.}
  \label{fig:gaugediag}
  \vspace{-2mm}
\end{figure}

Consider the process $gg \to \tilde{q}^*\tilde{q}$, for which the diagrams are 
shown in Fig.~\ref{fig:gaugediag}, where we wish the outgoing squarks to have 
masses $m_3$ and $m_4$ respectively. Due to the presence of external gluons, 
the Ward identity
\begin{equation} \label{eqn:ward}
  p_{1\mu}p_{2\nu}\me^{\mu\nu}(p_1,p_2) = 0
\end{equation}
where $p_i$ are the momenta of the gluons and $\me$ is the total amplitude, 
must be satisfied.

After replacing the external polarization vectors with their momenta the 
amplitudes are given by
\begin{subequations}
  \label{eqn:amplitudes}
  \begin{eqnarray}
    \me_t & = &-g_s^2\frac{\left(t-m_3^2\right)}{\left(t - m_t^2 \right)}
    \left(m_4^2 - t\right)
    t^b_{di} t^a_{ic},\label{eqn:ampt} \\
    \me_u  & = & -g_s^2\frac{\left(u-m_4^2\right)}{\left(u - m_u^2 \right)}
    \left(m_3^2 - u\right)
    t^a_{di} t^b_{ic}, \label{eqn:ampu} \\
    \me_s & = & -\frac{g_s^2}{2}\left(t - u\right)
    \left( t^b_{di} t^a_{ic} - t^a_{di} t^b_{ic}\right), \label{eqn:amps} \\
    \me_c & = & \frac{g_s^2}{2}s
    \left( t^a_{di} t^b_{ic} + t^b_{di} t^a_{ic}\right) \label{eqn:ampc},
  \end{eqnarray}
\end{subequations}
where $s$, $t$ and $u$ are the Mandelstam variables, $m_{t,u}$ are the $t$- and
$u$-channel masses and $t^a_{ij}$ are the $\mr{SU}(3)$ colour matrices. 
Equations~(\ref{eqn:ampt}) and~(\ref{eqn:ampu}) show that for any hope of 
achieving the correct cancellation we must set $m_t = m_3$ and $m_u = m_4$. 
This also shows why, even in the on-shell case, one must take care when using 
widths in scattering diagram calculations as these alone can give rise to  
violations of gauge invariance. The total amplitude saturated with the
gluon momenta for the $g\,g\to \tilde{q}^*\tilde{q}$ process is then
\begin{equation} \label{eqn:totamp}
  \frac{g_s^2}{2}\left(m_3^2 - m_4^2\right)\left[ t^b, t^a \right],
\end{equation}
so that Eq.~(\ref{eqn:ward}) is only satisfied if $m_3 = m_4$. 

\begin{figure}[t]
  \centering
  \includegraphics[angle=90,width=0.45\textwidth]{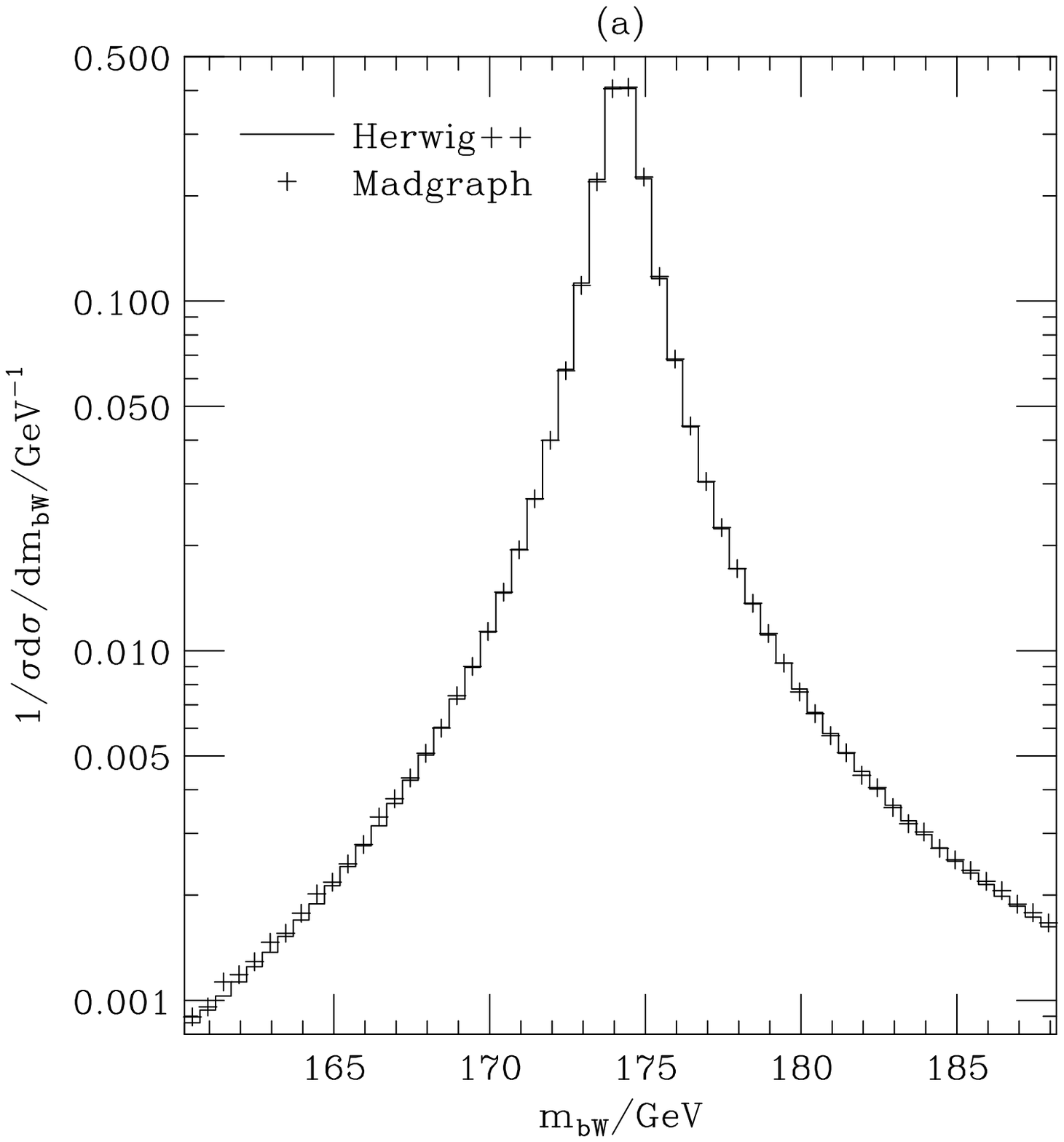}
  \includegraphics[angle=90,width=0.45\textwidth]{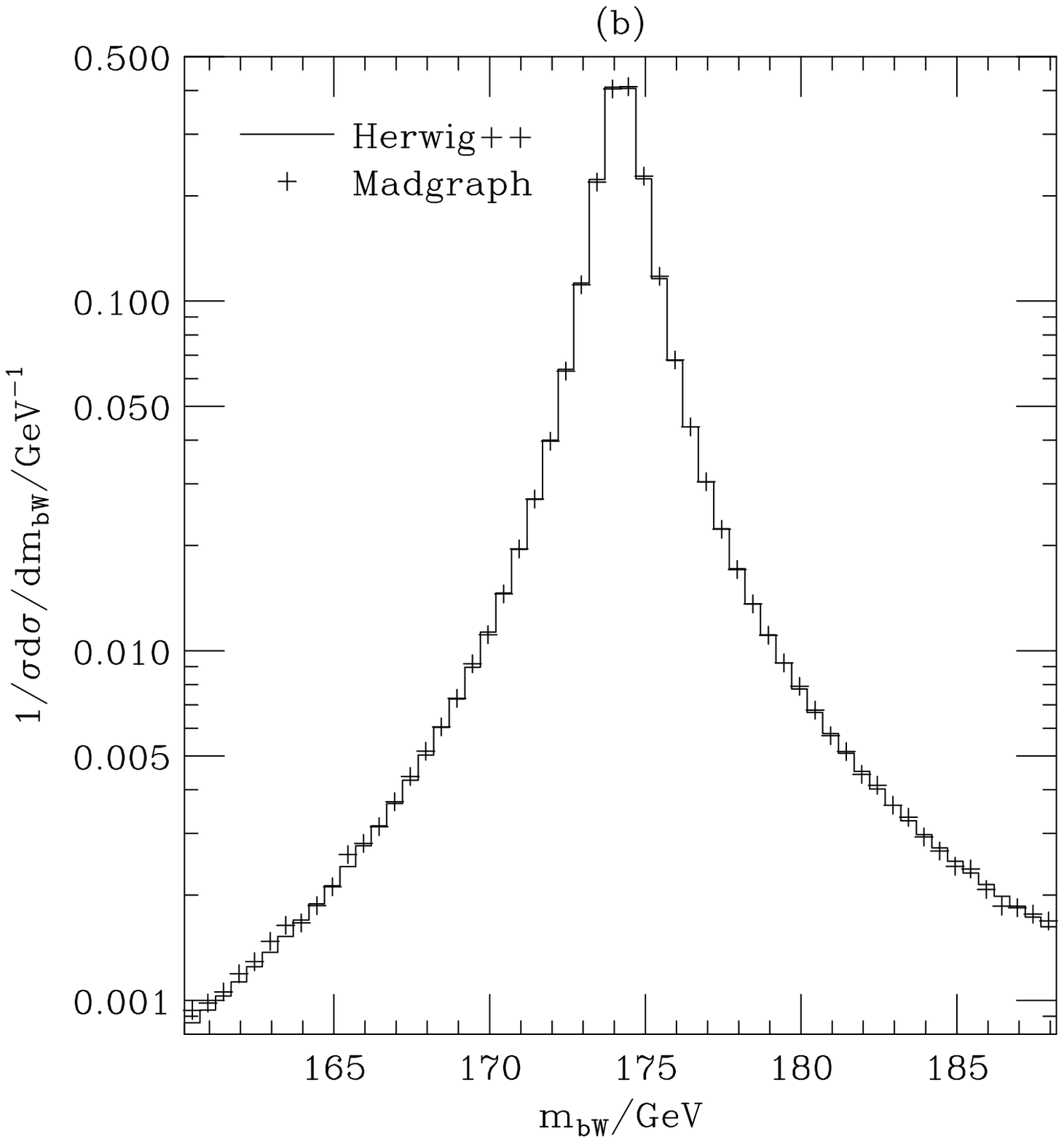}
  \caption{The top quark line shape for $m_t^{\mr{pole}}=174.20\mrgev$ 
    and $\Gamma_t(m_t^{\mr{pole}})=1.40\mrgev$ at the ILC with the results of
    \HWPP for $e^+e^- \to t\bar{t}\to b\bar{b}W^+W^-$ compared to
    the Madgraph calculation of $e^+e^-\to b\bar{b}W^+W^-$
    with the Madgraph result including (a) all diagrams containing a top quark
    line and (b) all diagrams excluding those containing a Higgs boson. In both
    cases the \HWPP result uses our off-shell treatment while the Madgraph 
    result includes all diagrams for the $2\to 4$ scattering process, including
    the non-resonant contribution.}
  \label{fig:leptopmass}
\end{figure}

This requirement means that, in general, we cannot use off-shell masses 
when calculating matrix elements since if we generate a process such as
that shown, we would violate gauge invariance. In our procedure the off-shell
masses are used when calculating the momenta of the outgoing particles involved 
in the hard interaction but are then rescaled, such that $m_3 = m_4$, for the matrix 
element calculations. To demonstrate the validity of this procedure we compare
the line shape of the top quark from \HWPP and Madgraph for the production
of a top quark at the ILC, the Tevatron and the LHC. In \HWPP the top quark 
width is computed using the full three-body matrix element whereas in the Madgraph
case just the two-body decay of the top quark is used, due to rapid
growth in the number of diagrams that are required. In all cases \HWPP generates
the $2\to 2$ production process for the $t\bar{t}$ pair followed by the three-body
decay of the top quark using the treatment of off-shell effects described in the
text. Madgraph was used to calculate the $2\to 4$ matrix element for the production
of $b\,\bar{b}\,W^-\,W^+$ including the non-resonant diagrams.

To ensure that the amplitudes generated by 
Madgraph~\cite{Maltoni:2002qb} were gauge invariant, the ``fudge-factor'' 
scheme~\cite{Beenakker:1996kt}
was used. This involves calculating the full amplitude without the inclusion of 
the width for any off-shell propagators and then multiplying the full amplitude, 
including non-resonant contributions, by
\begin{equation}
\label{eqn:fudgefactor}
\frac{p^2-M^2}{p^2-M^2 + iM\Gamma}
\end{equation}
for any propagator that can be on-shell, \ie for which it is possible for $p^2=M^2$
within the physically allowed phase space. This is the simplest approach that
ensures the amplitude is gauge invariant~\cite{Beenakker:1996kt}, although
it has the drawback that the non-resonant diagrams are affected. A more detailed
discussion of the issue of gauge-invariance when including non-resonant diagrams
can be found in~\cite{Beenakker:1996kt}.

\begin{figure}
  \centering
  \includegraphics[angle=90,width=0.45\textwidth]{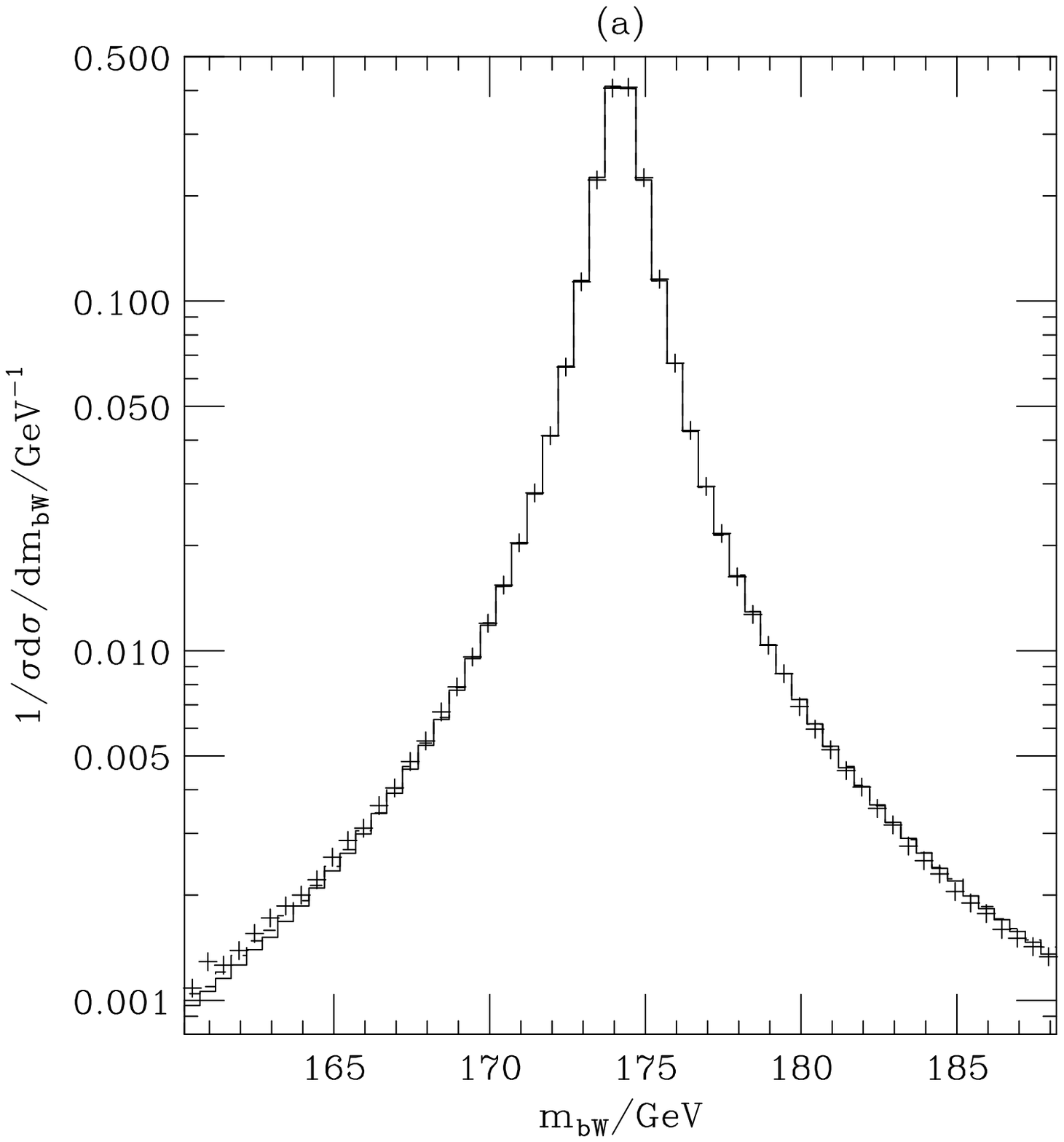}
  \includegraphics[angle=90,width=0.45\textwidth]{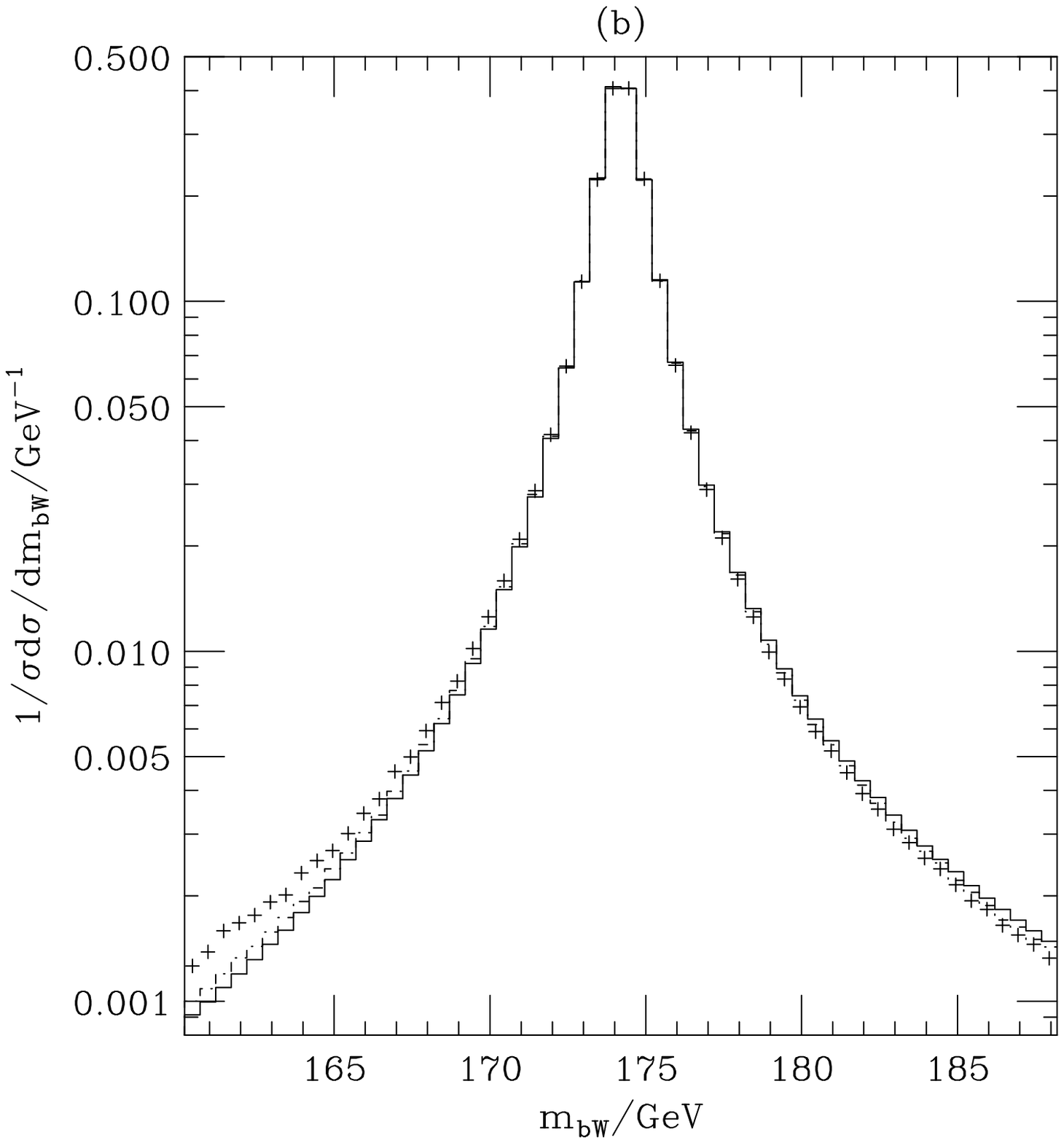}
  \caption{The top quark line shape for $m_t^{\mr{pole}}=174.2\mrgev$ and 
    $\Gamma_t(m_t^{\mr{pole}})=1.4\mrgev$ at (a) the Tevatron and (b) the LHC. The black 
    line denotes the results from \HWPP with the outgoing masses rescaled to
    their on-shell value, the dot-dash line denotes \HWPP with the outgoing
    masses rescaled to their average value and the crosses denote the Madgraph
    results. The \HWPP results were generated using the $2\to 2$ production
    process for $t\bar{t}$ followed by the three-body decay of the top quarks.
    The Madgraph results use the full matrix element for the production of
    $b\bar{b}W^+W^-$ to order $\alpha_S^2\alpha_W^2$, including all diagrams,
    both resonant and non-resonant diagrams, containing a top quark line.}
  \label{fig:hadrontopmass}
\end{figure}

For the ILC case, Fig.~\ref{fig:leptopmass}, 
the Madgraph result is shown for both the process including only the diagrams with
a top quark line and also the process including all electroweak diagrams, resonant and
non-resonant, excluding the Higgs. There is excellent agreement between 
our results and those performed with the full matrix element giving us 
confidence in our procedure. For the hadron colliders we must consider the rescaling since there will be
processes such as $gg\to t\bar{t}$ that will, as discussed above, violate
gauge invariance when we take the top quark off-shell. Here we will compare two
choices for the momenta rescaling, first rescaling such that the masses have
their on-shell value and second rescaling to the average value of the 
outgoing masses $(m_3 + m_4)/2$. The results for the Tevatron and the LHC
are shown in Fig.~\ref{fig:hadrontopmass}.
The Tevatron results are in excellent agreement with the matrix element
for both choices of rescaling and the LHC is good agreement except for the 
tail where there is a small deviation. It is clear that either choice for
the value of the rescaled mass gives good agreement with the matrix element
but in the LHC case choosing to rescale to the average value of the outgoing 
masses gives slightly better agreement with the full calculation.

\begin{figure}
  \centering {
  \includegraphics[angle=90,width=0.46\textwidth]{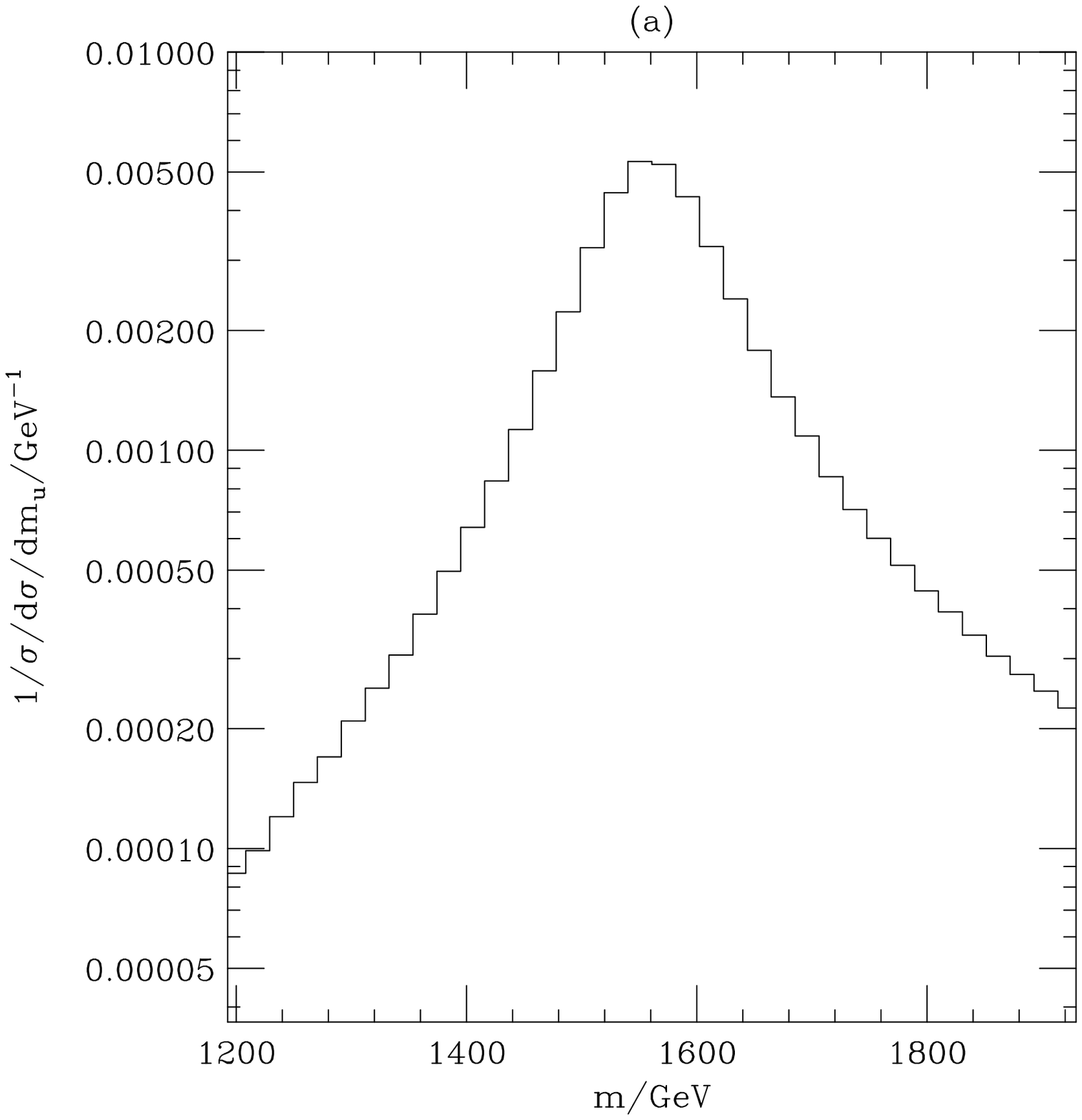}
  \includegraphics[angle=90,width=0.46\textwidth]{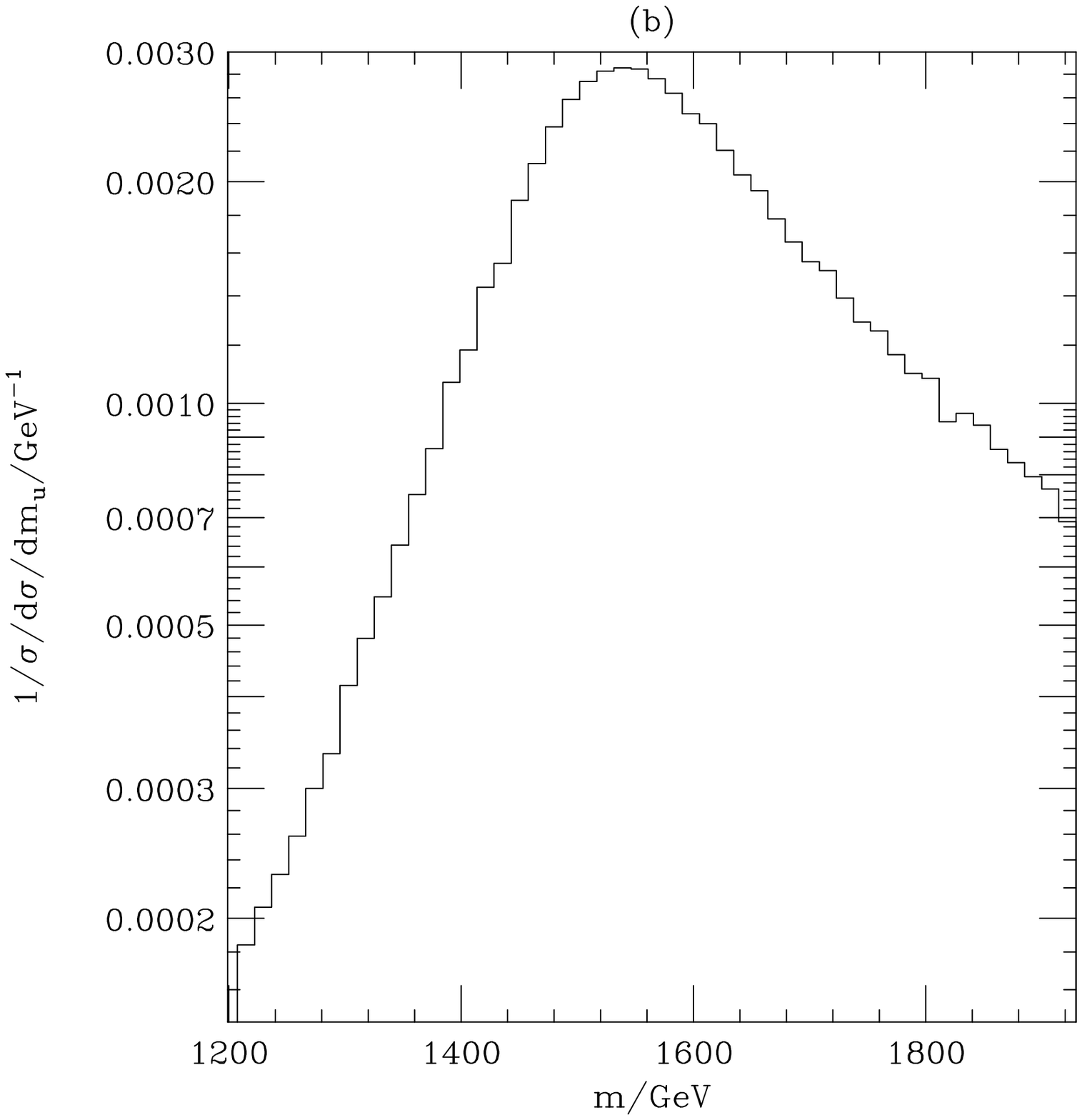}
  }
  \caption{The line shapes for (a) the SUSY partner to the $u$-type doublet field 
    and (b) the level-$1$ KK partner to the $u$-type doublet field.}
  \label{fig:ustarlineshape}
\end{figure}

Figure~\ref{fig:ustarlineshape} shows the mass distributions
for a left-handed up squark in the MSSM\footnote{It is technically incorrect 
to say that a scalar particle has a helicity state but the terminology allows 
for easier distinction between the partners of the $\mr{SU}(2)$ doublet and 
singlet quarks.} and the KK-partner of the doublet quark in the MUED model 
respectively. 
The mass spectrum for the MUED case is matched to the SUSY spectrum at SPS point 
2~\cite{Allanach:2002nj} where \mbox{$m_{\tilde{u}_L}=1560.97\mrgev$}, 
$\Gamma_{MSSM} = 70.22\mrgev$ and $\Gamma_{MUED} = 312.76\mrgev$.

This example is at the extreme of where this method should be applied since, 
especially in the MUED case, the width is large and in general there could be 
sizeable non-resonant contributions.

\subsection{Off-Shell Effects in Particle Decay}
\label{sec:particledecay}
Many new physics models have spectra that result in long
chains between the production of a resonance and a stable state. As mentioned 
previously
the simplest approach in dealing with these chains is via a series of on-shell
cascade decays. While this may be an appropriate approximation in some kinematic 
regions, in others, \ie when the decaying particle is close to threshold, the 
effects from the off-shell propagator must be taken into account. 

\begin{figure}[t]
  \centering{
    \includegraphics[angle=90,width=0.65\textwidth]{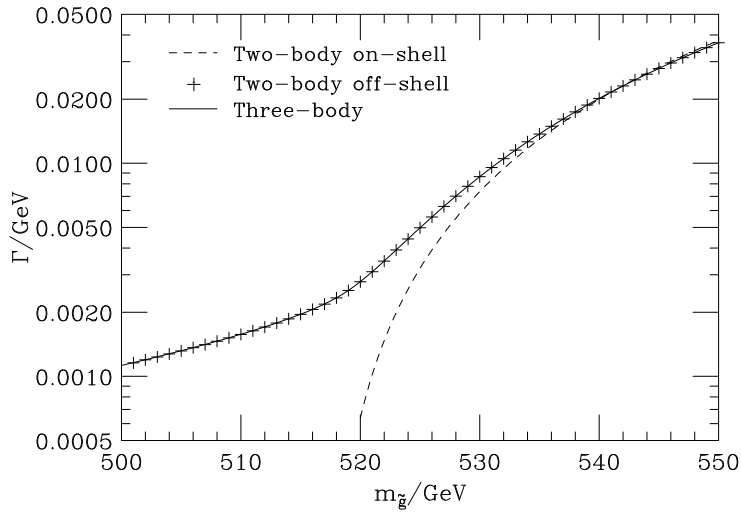}
  }
  \caption{The partial width for the decay mode 
    $\tilde{g}\to \bar{b}\,\tilde{b}_1 \to \tilde{\chi}_2^0\,b$ in the MSSM.}
  \label{fig:gtilb1}
\end{figure}

This can be achieved by including the weight factor from 
Eq.~(\ref{eqn:ofswgt}) in the calculation of the partial width of a selected decay 
mode. For example, consider the decay 
\mbox{$\tilde{g}\to \bar{b}\,\tilde{b}_1 \to \tilde{\chi}_2^0\,b\,\bar{b}$}, 
the partial width is
\begin{equation}
  \Gamma\left(\tilde{g}\to \tilde{\chi}_2^0\,b\,\bar{b}\right) = 
  \frac{1}{\pi}\int_{m^2_\mr{min}}^{m^2_\mr{max}} \mr{d}m^2
  \frac{m \Gamma\left(\tilde{b}_1 \to \tilde{\chi_2^0}\,b\right)}
       {\left(m^2 - M^2\right)^2 + M^2\Gamma(m)^2}
       \Gamma\left(\tilde{g} \to \bar{b}\,\tilde{b}_1\right),
\end{equation}
where the widths inside the integral are evaluated for the off-shell
mass $m$. The limits on the integration are determined by the on-shell width and are 
set such that the maximum deviation from the pole mass of $\tilde{b}_1$ is $5\Gamma$. 
As the intermediate particle is a scalar, the inclusion of the weight
factor should give exact agreement with the full three-body calculation
providing the integral is performed over the same phase space. Fig~\ref{fig:gtilb1} 
demonstrates this for SPS point 1a where the three-body phase-space is
restricted to the same as the two-body case.
The spectrum was produced using \textsf{SPheno} 2.2.3~\cite{Porod:2003um} where 
$m_{\tilde{b}_1}=515.27\mrgev$, $\Gamma(m_{\tilde{b}_1})=3.83\mrgev$ and 
$m_{\tilde{\chi}_2^0}=180.58\mrgev$. The 
mass of the $b$-quark is $m_b=4.20\mrgev$, which is the default value in \HWPP.
The on-shell result is also included for reference.

The agreement between the full matrix element calculation and our results show
that the approximation is valid.

\section{Examples}
\label{sec:examples}
Here we present a range of processes in the MSSM and the MUED model demonstrating the 
consistency of the inclusion of off-shell effects in \HWPP. In the decay
examples the comparison is always to the full three-body result, which was
included in \HWPP as a modification to the current released version and will
be included in a future version along with the simulation of finite width effects.

\subsection{Decay via an Off-Shell Fermion}
\begin{figure}
  \centering{
    \includegraphics[angle=90,width=0.65\textwidth]{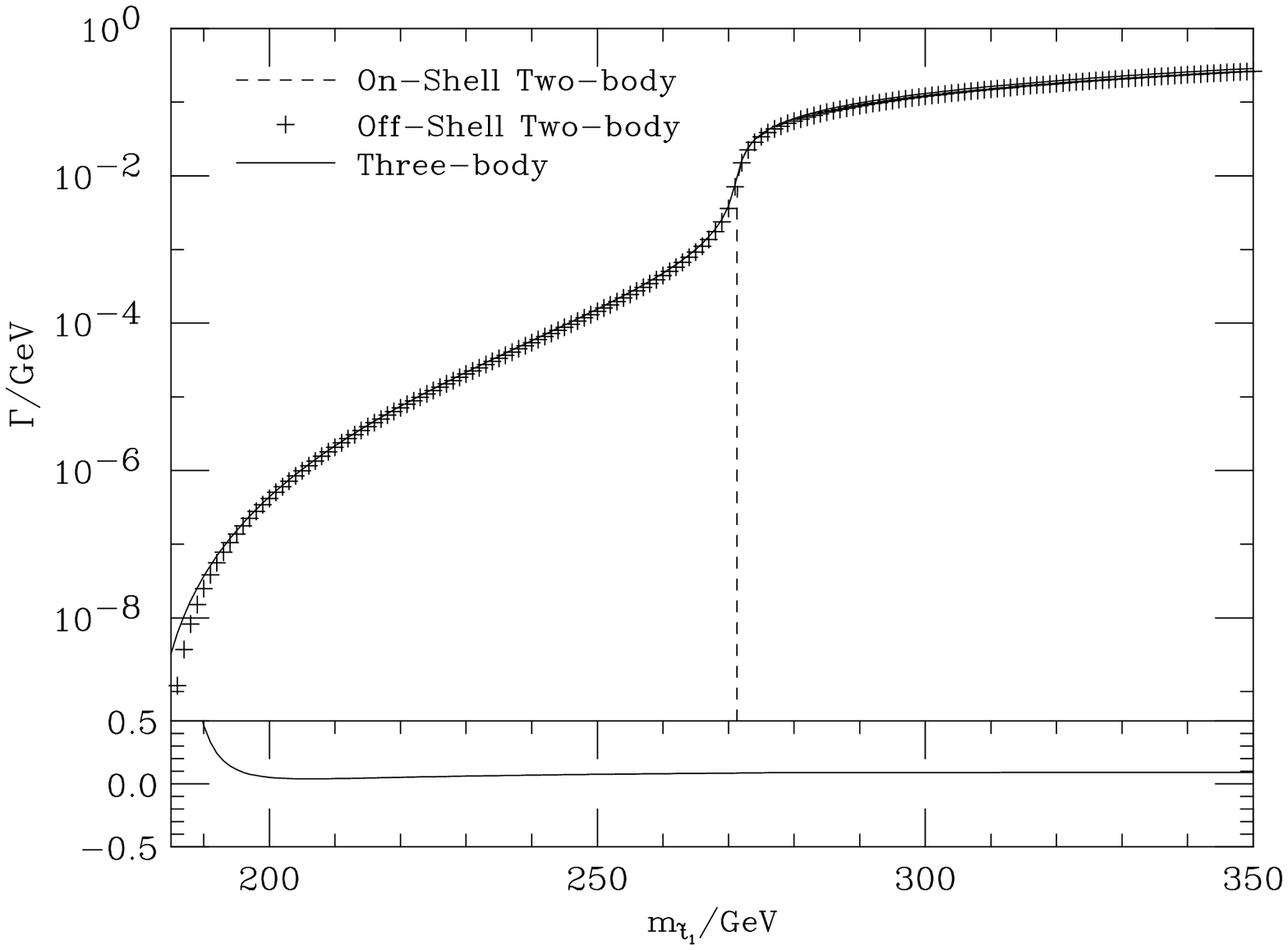}
  }
  \caption{The partial width for the decay mode 
    $\tilde{t}_1\to \tilde{\chi}_1^0\, t \to W^+\,b\,\tilde{\chi}_1^0$. The lower panel gives 
  the value of $\left(\Gamma_{\mr{three}}/\Gamma_{\mr{off}}-1\right)$.}
  \label{fig:stopofs}
\end{figure}

A possible two-body decay of the $\tilde{t}_1$ squark in the MSSM is 
$\tilde{t}_1 \to \tilde{\chi}_1^0 t$. If $m_{\tilde{t}_1} \approx  
m_{\tilde{\chi}_1^0} + m_t$ then the effect of the width of the top quark must be 
considered. We choose the decay mode $\tilde{t}_1 \to \tilde{\chi}_1^0\, t
\to \tilde{\chi}_1^0\, W^+b$ at SPS point 1a~\cite{Allanach:2002nj}
where $m_{\tilde{\chi}_1^0}=97.04\mrgev$ with $m_t=174.20\mrgev$,
$m_W=80.40\mrgev$ and $m_b = 4.20\mrgev$. The threshold values for the on-shell 
two- and three-body decays of the $\tilde{t}_1$ are $271.24\mrgev$ and 
$181.64\mrgev$ respectively. Figure~\ref{fig:stopofs} shows partial width of 
the $\tilde{t}_1$ as a function of its mass for the three-body, two-body off-shell 
and two-body on-shell results.

Unlike the gluino decay example in Sect.~\ref{sec:particledecay} the $W^\pm$
boson, like the top quark, has a measured decay width and this should be treated
properly. In the example shown in Fig.~\ref{fig:stopofs} the running width for
the top quark is calculated from its full three-body matrix element to a $b$-quark 
and a pair of light fermions which includes the full effects of the $W^\pm$ width. The 
agreement between the two-body off-shell and three-body results shows that this is
a valid approximation to use. Also, despite the extra factor of $(\sla{p}+m)$ in the 
numerator of the fermion propagator, there is still good agreement between the 
full three-body result and the two-body result with weight factor even though
the factor does not attempt to include this. This is due to the numerator factor being
largely responsible for propagating spin information rather than altering the 
kinematics.

\subsection{Decay via an Off-Shell Gauge Boson}
In the MSSM there is a coupling between the $Z^0$ boson and the gaugino sector
allowing for a three-body decay of the second neutralino to a pair of
light fermions via an intermediate $Z^0$ boson. The presence of a spin-$1$
rather than a spin-$0$ particle alters the form of the partial width as the 
decay is now $p$-wave and not $s$-wave, as in Fig.~\ref{fig:gtilb1}.
To illustrate that the weight formula works just as well in this situation 
we choose the decay chain $\tilde{\chi}_2^0 \to \tilde{\chi}_1^0\,Z^0\to \tilde{\chi}_1^0\,b\,\bar{b}$
at SPS point 1a where $m_{\tilde{\chi}_1^0}=97.04\mrgev$ with
$M_Z=91.19\mrgev$ and $m_b=4.20\mrgev$.

\begin{figure}
  \centering{
    \includegraphics[angle=90,width=0.65\textwidth]{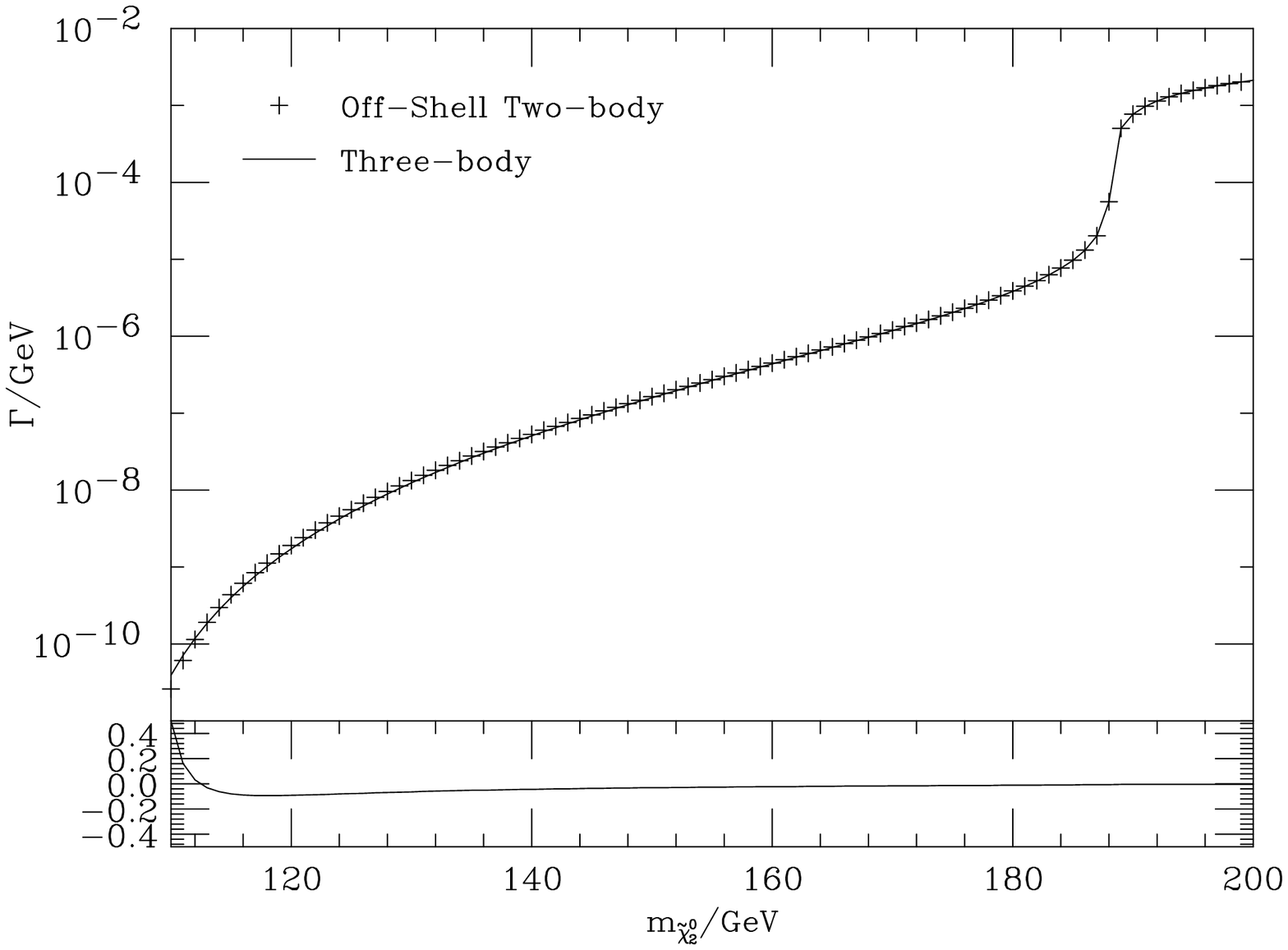}
  }
  \caption{The partial width for 
    $\tilde{\chi}_2^0 \to \tilde{\chi}_1^0\,Z^0 \to b\,\bar{b}\,\tilde{\chi}_1^0$  
    in the MSSM. The lower panel gives 
  the value of $\left(\Gamma_{\mr{three}}/\Gamma_{\mr{off}}-1\right)$.}
  \label{fig:chi2bb}
\end{figure}

\begin{figure}
  \centering{
    \includegraphics[angle=90,width=0.65\textwidth]{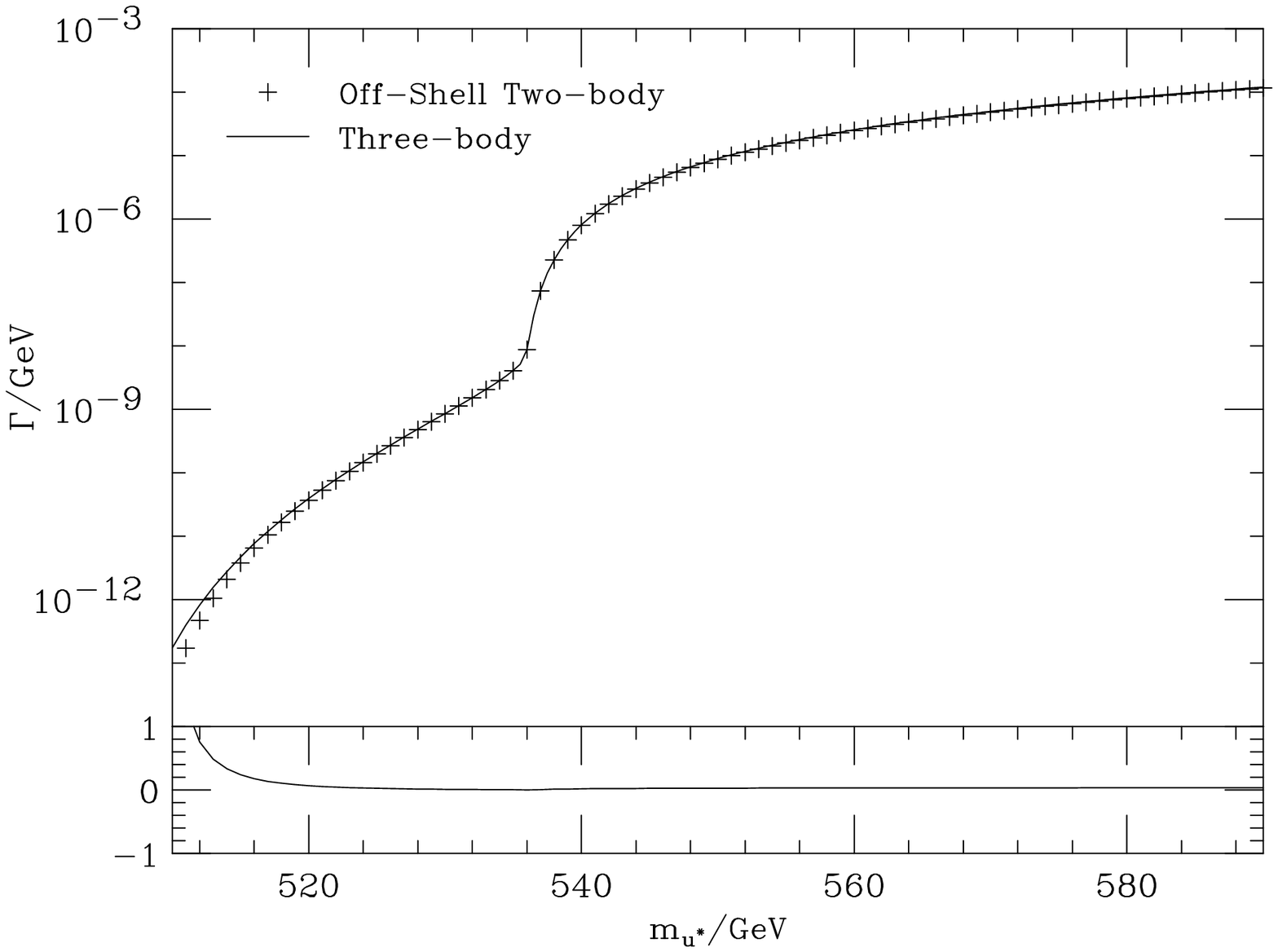}
  }
  \caption{The partial width for $u^\bullet \to u\,e_1^{\circ -}e^+$
      in the MUED model. The lower panel gives 
  the value of $\left(\Gamma_{\mr{three}}/\Gamma_{\mr{off}}-1\right)$.}
  \label{fig:ueduLee}
\end{figure}

Figure~\ref{fig:chi2bb} shows the results for the above decay and demonstrates
that there is good agreement between the full three-body result and
the two-body approximation for an intermediate vector particle. Another
example of a possible $p$-wave decay is $u^\bullet \to u\,e_1^{\circ -}e^+$ in the
MUED model where the intermediate particle is the level-$1$ KK-$Z^0$ boson. For
parameter values $R^{-1}=500\mrgev$ and $\Lambda R=20$ the relevant masses
are $M_{Z_1^0}=535.81\mrgev$ and $M_{e^\circ_1}=504.25\mrgev$. The partial
width is shown in Fig.~\ref{fig:ueduLee}, again with both the three-body result
and two-body via an off-shell $Z^0_1$.

\section{Off-Shell Cross Sections}
In the narrow width approximation, the cross section for a particular 
final state is computed by taking the on-shell production cross section to
an intermediate resonance and multiplying by the branching fraction
to the final state of interest. If the on-shell mass of the resonance is
close the threshold for the decay into the final state then the narrow width
approximation is invalid and one should calculate the full matrix element. As described
in Sect.~\ref{sec:offshell} and demonstrated in Sect.~\ref{sec:examples}
we can include a weight factor in production and decay to simulate
off-shell behaviour. In the case of calculating cross sections for specific 
processes, this amounts to including the effect of propagator widths in the
Monte Carlo estimate of such a quantity. It is important to note, however, that
a general purpose event generator that starts from a $2 \to 2$ hard scattering
and then perturbatively decays the produced resonances can never include
non-resonant contributions. This is a fundamental limit of the approximations
used to generate the events. Nevertheless, a good approximation can still
be achieved providing one uses the simulations with care\footnote{In some specific
cases non-resonant effects can be modelled by using a modified form of 
Eq.~(\ref{eqn:ofswgt}), for example the Higgs lineshape~\cite{Seymour:1995qg}.}. 

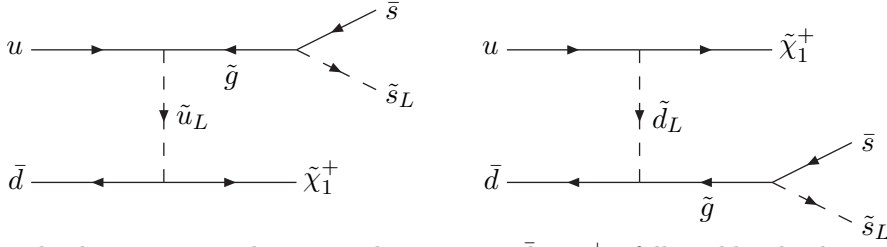
\begin{figure}
  \centering{
    \begin{picture}(200,90)(70,10)
      \ArrowLine(10,70)(60,70)          \Text(7,71)[r]{$u$}
      \ArrowLine(110,70)(60,70)         \Text(83,60)[l]{$\tilde{g}$}
      \DashArrowLine(60,70)(60,20){5}   \Text(65,45)[l]{$\tilde{u}_L$}
      \ArrowLine(60,20)(10,20)          \Text(7,22)[r]{$\bar{d}$}
      \ArrowLine(60,20)(110,20)         \Text(113,22)[l]{$\tilde{\chi}_1^+$}
      \ArrowLine(140,85)(110,70)        \Text(144,85)[l]{$\bar{s}$}
      \DashArrowLine(110,70)(140,55){5} \Text(144,53)[l]{$\tilde{s}_L$}

      \SetOffset(180,0)

      \ArrowLine(10,70)(60,70)           \Text(7,71)[r]{$u$}		  
      \ArrowLine(60,70)(110,70)		 \Text(113,72)[l]{$\tilde{\chi}_1^+$}
      \DashArrowLine(60,70)(60,20){5}	 \Text(65,45)[l]{$\tilde{d}_L$}	  
      \ArrowLine(60,20)(10,20)		 \Text(7,22)[r]{$\bar{d}$}		  
      \ArrowLine(110,20)(60,20)		 \Text(83,10)[l]{$\tilde{g}$}
      \ArrowLine(140,35)(110,20)         \Text(144,35)[l]{$\bar{s}$}    
      \DashArrowLine(110,20)(140,5){5}	 \Text(144,3)[l]{$\tilde{s}_L$}
    \end{picture}
  }
  \caption{The diagrams contributing to the process
    $u\,\bar{d}\to \tilde{\chi}_1^+\, \tilde{g}$ followed by the decay of the
  gluino to a strange quark and a left-handed strange squark.}
\label{fig:xsecdiagrams}
\end{figure}

\begin{figure}
  \centering{
    \includegraphics[angle=90,scale=0.6]{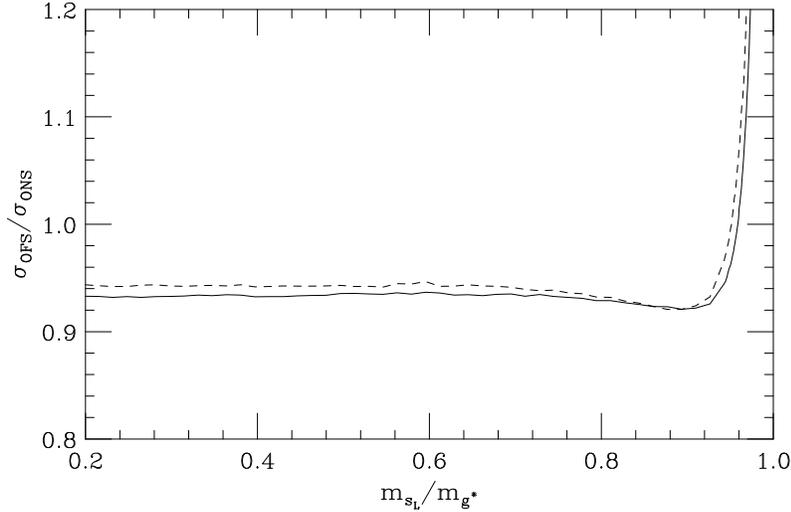}
  }
  \vspace{-3mm}
  \caption{The ratio of the off-shell and on-shell
    cross section for the process $u\,\bar{d}\to \tilde{\chi}_1^+\,
    \tilde{s}_L\,\bar{s}$~(black line) and its MUED counterpart~(dashed line).}
  \label{fig:udxsecplot}
\end{figure}

An example of a process that has no non-resonant contributions is the
production of a strange squark via 
$u\,\bar{d}\to \tilde{\chi}_1^+\, \tilde{g}\to\tilde{\chi}_1^+\,\tilde{s}_L\,\bar{s}$, 
the diagrams
for which are shown in Fig.~\ref{fig:xsecdiagrams}. The results for the 
ratio of the off-shell to the on-shell cross section as a function of the 
strange squark mass are shown in Fig.~\ref{fig:udxsecplot} for SPS point 1a. The 
ratio is constant, with the off-shell result smaller due to the integration limits 
no longer being taken to infinity, until $m_{\tilde{s}_L}\approx 0.8 m_{\tilde{g}}$ where we are 
in the threshold region for the decay of the gluino. 
The sudden steep rise as the mass ratio approaches unity is due to the 
on-shell cross section going to zero at threshold. 

There is a counterpart process to that in Fig.~\ref{fig:xsecdiagrams} for MUED,
where the $\tilde{\chi}_1^+$ is replaced by the $W^+_1$ boson, the $\tilde{g}$
by the $g_1$ and $\tilde{s}_L$ by the $s^\bullet_1$. The ratio of the on-shell
to the off-shell cross section for this process is also shown in 
Fig.~\ref{fig:udxsecplot} where the masses for the MUED particles have been
matched to SUSY spectrum to give a fair comparison. It is apparent here that
the spins of the underlying model play only a small role in determining the
value of this ratio as the results are similar and while the absolute values
of the cross sections may differ greatly, taking the colour octet object off-shell 
affects only the kinematics.

\section{Merging Two- and Three-Body Decays}
In Sect.~\ref{sec:examples} we demonstrated the accuracy of including
an off-shell weight factor by comparison with the full three-body 
matrix element for a variety of processes. For each process considered the
width was plotted over the entire kinematic range, rather than restricting to
the region where the decay would be applicable, to give a full comparison. In
a real simulation there is a choice over which point we should change between 
using a two- and three-body decay of the particle. If both decays were 
treated on-shell then the point would simply be the threshold of the 
two-body decay but when including of off-shell effects for the two-body 
decay, the choice is not so simple.

\begin{figure}
  \centering
  \includegraphics[angle=90,scale=0.4]{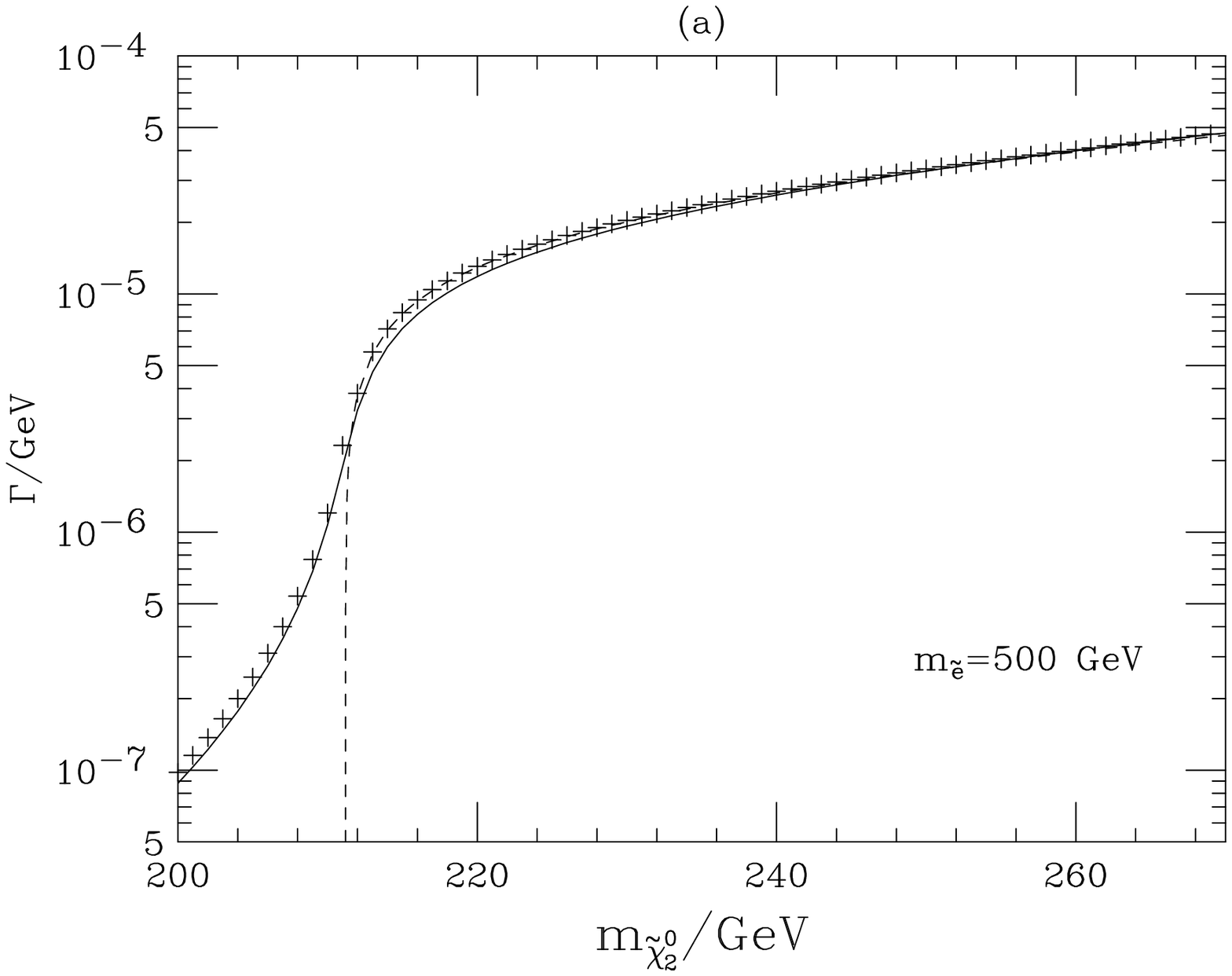}
  \includegraphics[angle=90,scale=0.4]{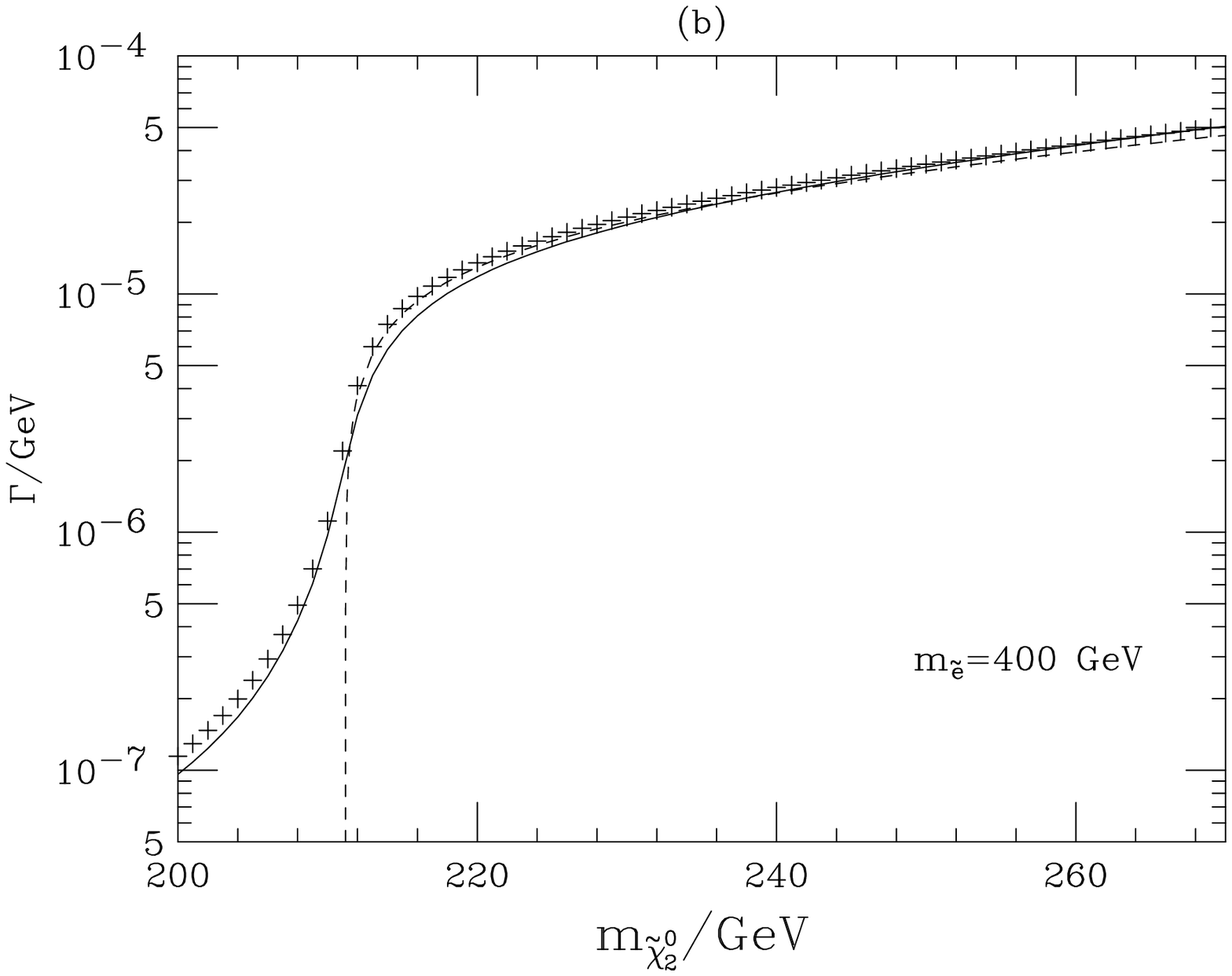}\\
  \includegraphics[angle=90,scale=0.4]{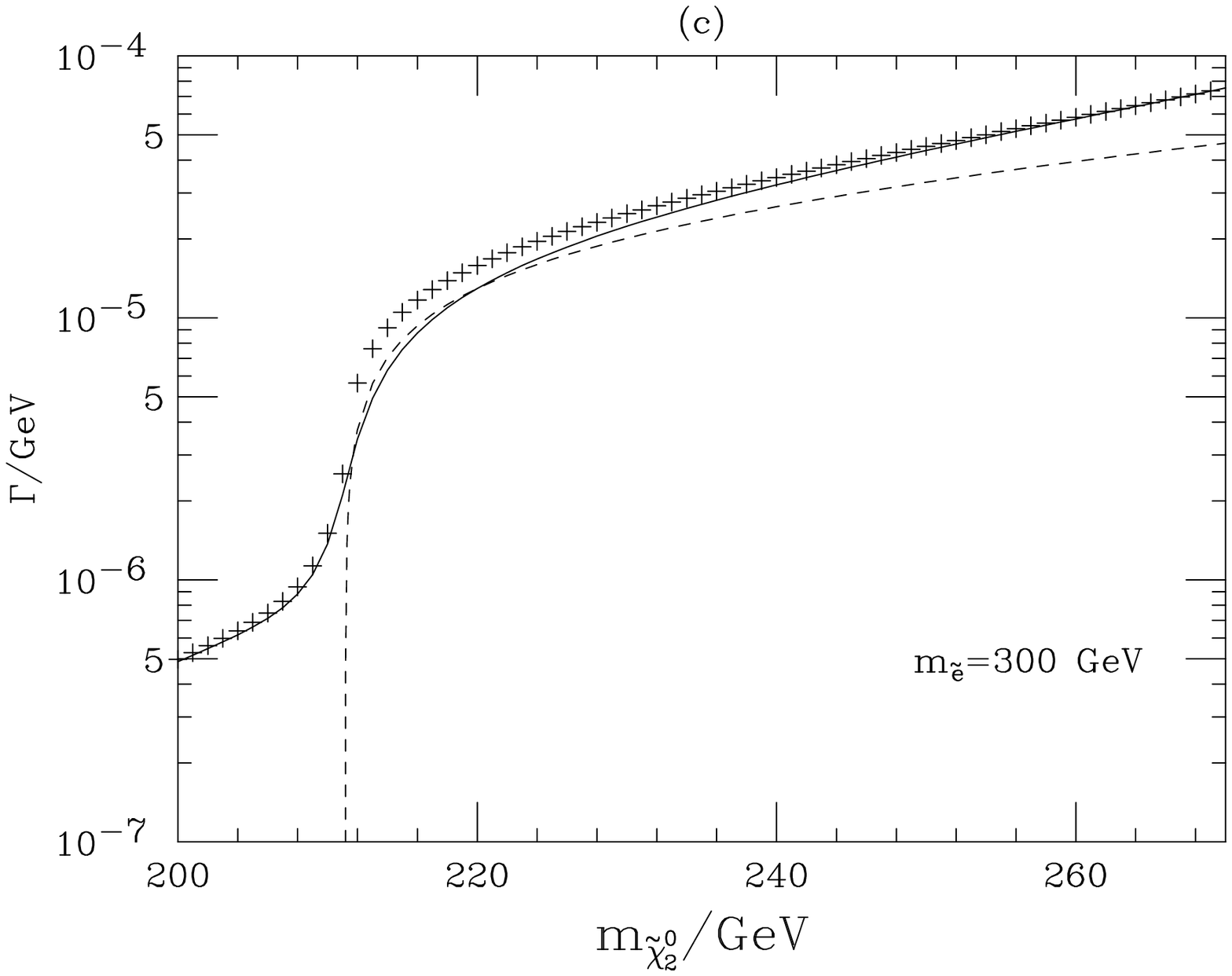}
  \includegraphics[angle=90,scale=0.4]{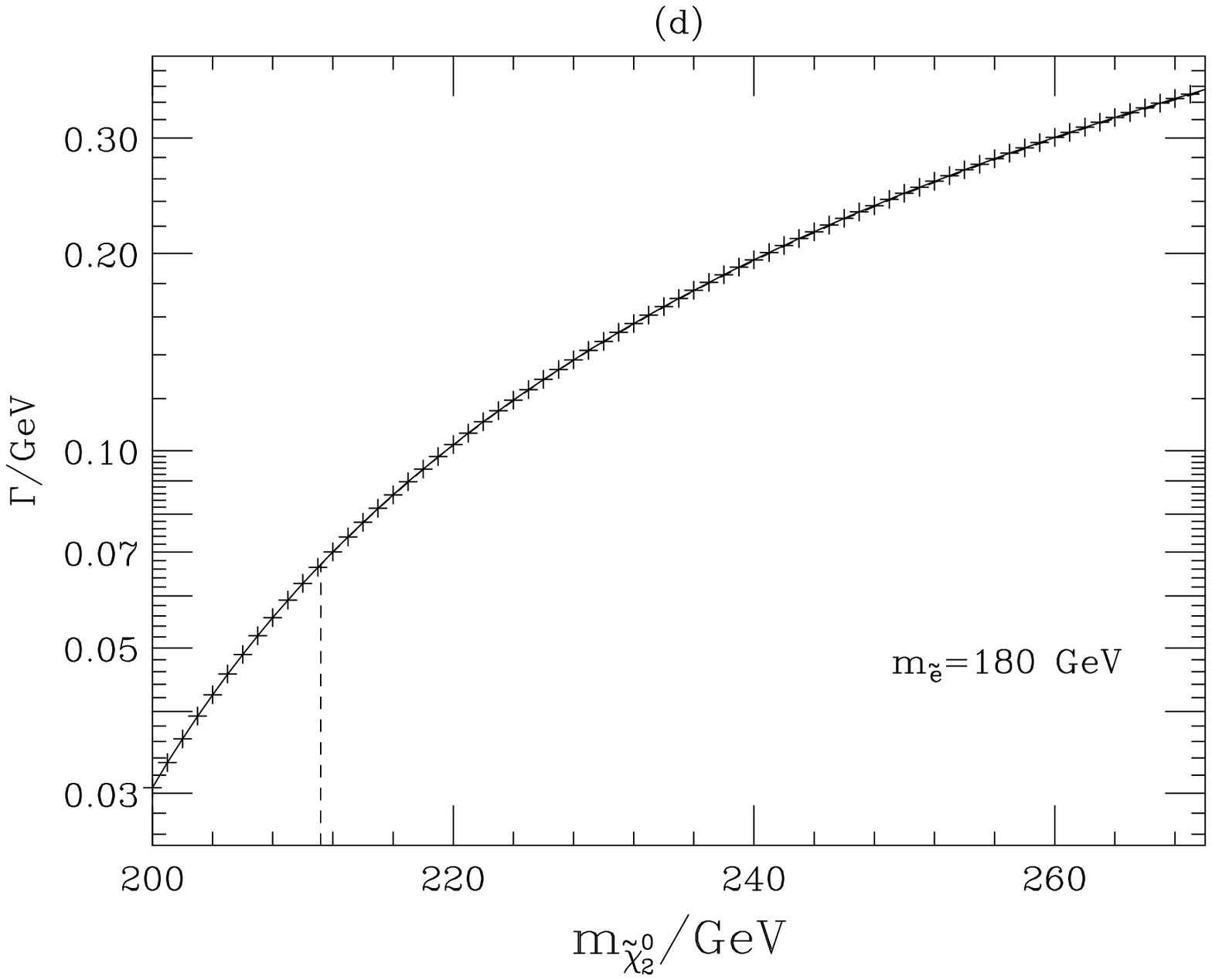}
  \caption{The partial width for the decay 
    $\tilde{\chi}_2^0 \to \tilde{\chi}_1^0\,e^+\,e^-$ where the selectron masses
    are indicated on the plot.
    The solid line is the full three-body partial width and the crosses are 
    for the three-body decay with the $Z^0$ diagram removed plus 
    $\Gamma(\tilde{\chi}^0_2\to\tilde{\chi}_1^0\,Z^0)\times BR(Z^0\to e^+\,e^-)$.
    The dashed line shows the two-body on-shell cascade.}
  \label{fig:crossover}
\end{figure}

Here we use the three-body decay $\tilde{\chi}_2^0 \to 
\tilde{\chi}_1^0\,e^+\,e^-$ in the MSSM to study this effect. The full three-body
decay is mediated by a $Z^0$ boson and both the left- and right-handed selectron,
giving an interference between the different channels. If the decay occurs as a 
series of cascades with a weight factor then these interference effects will be 
neglected. To judge the extent of the interference we compare the
partial width for the decay via the full three-body matrix element and the 
three-body matrix element with the $Z^0$ diagram removed and performed as a 
cascade decay.

Figure~\ref{fig:crossover} shows the results for a range of selectron masses, 
where both
$\tilde{e}_L$ and $\tilde{e}_R$ are degenerate, with $M_{\tilde{\chi}_1^0}=120.00\mrgev$ 
and $M_Z=91.19\mrgev$. For a sufficiently large selectron mass there is good agreement 
between the two methods as there is only a small interference with the $Z^0$ boson 
diagram. However, as the mass is lowered so that the decay of the 
$\tilde{\chi}_2^0$ through the selectron mode becomes closer to being on-shell, 
the interference effects, particularly just above the $Z^0$ threshold, 
become significant 
and the full three-body calculation is necessary in this region. For the final
case, Fig.~\ref{fig:crossover}c, where the selectron modes are on-shell for the 
whole range there is quite different behaviour. The partial width now smoothly
passes over the $Z^0$ threshold and there is exact agreement with the full 
three-body result indicating that there is very little or no interference.
Given these results it seems reasonable to use the threshold of the on-shell 
two-body decay as the point where the change from a three-body to a two-body decay 
with weight factor occurs.

\section{Summary}
Given that we do not know what type of new physics will be discovered at future
colliders, it is necessary to have access to accurate simulations of these
new models. An important consideration when studying new physics scenarios within
a Monte Carlo event generator is the simulation of off-shell effects. Here
we have demonstrated a consistent algorithm for their inclusion in a 
general-purpose event generator using a variety of processes from the 
Minimal Supersymmetric Standard and Minimal Universal Extra Dimension Models.
The approach described here has been implemented in the \HWPP program and will
be incorporated in to a future release.

\acknowledgments
We would like to thank our collaborators on the \HWPP project and
the organisers and participants of the Les Houches 2007 workshop for many 
useful comments.
This work was supported in part by the Science and Technology 
Facilities Council and the European Union Marie Curie Research Training
Network MCnet under contract MRTN-CT-2006-035606.

\appendix

\section{Derivation of Weight Factor}
\label{app:wgtderiv}

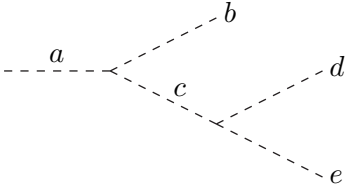
\begin{figure}
\centering{
  \begin{picture}(150,90)
    \DashLine(10,60)(50,60){3} \Text(27,66)[l]{$a$} 
    \DashLine(50,60)(90,80){3} \Text(93,82.5)[l]{$b$}
    \DashLine(50,60)(90,40){3} \Text(74,53)[l]{$c$}
    \DashLine(90,40)(130,60){3} \Text(133,62)[l]{$d$}
    \DashLine(90,40)(130,20){3} \Text(133,20)[l]{$e$}
  \end{picture}
}
\caption{A three-body decay consisting entirely of scalar particles where
the external particles $a$,$\,b$,$\,d$ and $e$ are all on-shell.}
\label{fig:wgtdiag}
\end{figure}

The weight factor introduced in Eq.~(\ref{eqn:ofswgt}) can be derived by 
considering a three-body decay that consists entirely of scalar particles.
Using the notation in Fig.~\ref{fig:wgtdiag}, the decay rate is 
given by
\begin{equation}
\label{eqn:threebody}
  \Gamma(a\to b,d,e) = \frac{\left(2\pi\right)^4}{2m_a}
\int \mr{d}\phi_3\left(p_a;p_b,p_d,p_e\right)
  \sum_{i=1}^n\mee{3_i}
\end{equation}
where $\mr{d}\phi_3$ is the three-body phase-space and 
$\mee{3}$ is the spin-averaged matrix element. The phase space
can be written recursively as~\cite{Yao:2006px,Jackson:1964zd}
\begin{equation}
  \label{eqn:phasedeco}
  \mr{d}\phi_3\left(p_a;p_b,p_d,p_e\right) 
  = \mr{d}\phi_2\left(p_a;p_b,q\right)\mr{d}\phi_2\left(q;p_d,p_e\right)
  \left(2\pi\right)^3\mr{d}q^2,
\end{equation}
where $\mr{d}\phi_2$ is a two-body phase-space factor and $q$ is the momentum of 
the intermediate.

For the matrix element assume that the intermediate particle of mass $M$ 
has $n$ two-body decay modes so that
\begin{equation}
  \label{eqn:medeco}
  \sum_{i=1}^n\mee{3_i} = \frac{\mee{a}}{\invPA{q}{M}}
  \sum_{i=1}^n \mee{c_i},
\end{equation}
where $\mee{a}$ and $\mee{c_i}$ are the two-body spin-averaged matrix 
elements and $\Gamma(q)$ is the width of the intermediate at scale $q$. Substituting 
Eqns.~(\ref{eqn:phasedeco},\ref{eqn:medeco}) into Eq.~(\ref{eqn:threebody}) gives
\begin{equation}
  \label{eqn:threetwo}
  \Gamma(a\to b,d,e) =\frac{\left(2\pi\right)^3}{m_a} 
  \int\mr{d}q^2 \frac{q}{\invPA{q}{M}}
  \int \mr{d}\phi_2 \mee{a}
  \int \mr{d}\phi_2 \frac{\left(2\pi\right)^4}{2q}
  \sum_{i=1}^n \mee{c_i}.
\end{equation} 
The third integral in Eq.~(\ref{eqn:threetwo})
can be recognised as $\Gamma(c\to d,e)$ using
\begin{equation}
  \Gamma(a\to b,c)= \frac{\left(2\pi\right)^4}{2m_a} \int \mee{a}\mr{d}\phi_2
\end{equation}
and the second as $\Gamma(a\to b,c)$, with the intermediate at scale $q$, giving
\begin{equation}
  \Gamma(a\to b,d,e) = \frac{1}{\pi}
  \int \mr{d}q^2\, \Gamma_a 
  \frac{q\Gamma_c(q)}{\left(q^2-M^2\right)^2 + q^2\Gamma_c^2(q)}.
\end{equation}
The weight factor is then identified as 
\begin{equation}
  w = \frac{1}{\pi}\int \mr{d}q^2\frac{q\Gamma(q)}{\invPA{q}{M}}.
\end{equation}
The case for production followed by decay follows similar arguments
and the same factor is found.

It should be noted from the discussion in Ref.~\cite{Seymour:1995qg} that
use of the running width in the propagator of Eqn.~(\ref{eqn:threetwo})
is only valid if $\Gamma(q)\sim q$ for large $q$. If $\Gamma(q)$ were to grow 
faster than this then the extra terms are dominant and
the propagator becomes of $\mr{O}(1/\alpha)$, the coupling of the $c\to d,e$
decay. If $\Gamma(q)$ grows linearly with $q$ the extra terms are just 
unenhanced higher order corrections.

\bibliographystyle{JHEP}
\bibliography{offshell}

\providecommand{\href}[2]{#2}\begingroup\raggedright\begin{thebibliography}{10}

\bibitem{Maltoni:2002qb}
F.~Maltoni and T.~Stelzer, {\it {MadEvent: Automatic Event Generation with
  MadGraph}},  {\em JHEP} {\bf 02} (2003) 027,
  [\href{http://xxx.lanl.gov/abs/hep-ph/0208156}{{\tt hep-ph/0208156}}].

\bibitem{Pukhov:1999gg}
A.~Pukhov {\em et.~al.}, {\it {CompHEP: A package for evaluation of Feynman
  diagrams and integration over multi-particle phase space. User's manual for
  version 33}},  \href{http://xxx.lanl.gov/abs/hep-ph/9908288}{{\tt
  hep-ph/9908288}}.

\bibitem{Pukhov:2004ca}
A.~Pukhov, {\it {CalcHEP 3.2: MSSM, structure functions, event generation,
  batchs, and generation of matrix elements for other packages}},
  \href{http://xxx.lanl.gov/abs/hep-ph/0412191}{{\tt hep-ph/0412191}}.

\bibitem{Kilian:2001qz}
W.~Kilian, {\it {WHIZARD 1.0: A generic Monte Carlo integration and Event
  Generation package for multi-particle processes. Manual}}, .
  LC-TOOL-2001-039.

\bibitem{Moretti:2001zz}
M.~Moretti, T.~Ohl, and J.~Reuter, {\it {O'Mega: An Optimizing Matrix Element
  Generator}},  \href{http://xxx.lanl.gov/abs/hep-ph/0102195}{{\tt
  hep-ph/0102195}}.

\bibitem{Krauss:2001iv}
F.~Krauss, R.~Kuhn, and G.~Soff, {\it {AMEGIC++ 1.0: A Matrix Element Generator
  In C++}},  {\em JHEP} {\bf 02} (2002) 044,
  [\href{http://xxx.lanl.gov/abs/hep-ph/0109036}{{\tt hep-ph/0109036}}].

\bibitem{Bahr:2008pv}
M.~Bahr {\em et.~al.}, {\it {Herwig++ Physics and Manual}},
  \href{http://xxx.lanl.gov/abs/0803.0883}{{\tt 0803.0883}}.

\bibitem{Corcella:2000bw}
G.~Corcella {\em et.~al.}, {\it {HERWIG 6: An event generator for hadron
  emission reactions with interfering gluons (including supersymmetric
  processes)}},  {\em JHEP} {\bf 01} (2001) 010,
  [\href{http://xxx.lanl.gov/abs/hep-ph/0011363}{{\tt hep-ph/0011363}}].

\bibitem{Moretti:2002eu}
S.~Moretti, K.~Odagiri, P.~Richardson, M.~H. Seymour, and B.~R. Webber, {\it
  {Implementation of supersymmetric processes in the HERWIG event generator}},
  {\em JHEP} {\bf 04} (2002) 028,
  [\href{http://xxx.lanl.gov/abs/hep-ph/0204123}{{\tt hep-ph/0204123}}].

\bibitem{Sjostrand:2006za}
T.~Sjostrand, S.~Mrenna, and P.~Skands, {\it {PYTHIA 6.4 Physics and Manual}},
  {\em JHEP} {\bf 05} (2006) 026,
  [\href{http://xxx.lanl.gov/abs/hep-ph/0603175}{{\tt hep-ph/0603175}}].

\bibitem{Sjostrand:2007gs}
T.~Sjostrand, S.~Mrenna, and P.~Skands, {\it {A Brief Introduction to PYTHIA
  8.1}},  \href{http://xxx.lanl.gov/abs/0710.3820}{{\tt arXiv:0710.3820}}.

\bibitem{Gleisberg:2003xi}
T.~Gleisberg {\em et.~al.}, {\it {SHERPA 1.$\alpha$, A Proof-of-Concept
  Version}},  {\em JHEP} {\bf 02} (2004) 056,
  [\href{http://xxx.lanl.gov/abs/hep-ph/0311263}{{\tt hep-ph/0311263}}].

\bibitem{Knowles:1988vs}
I.~G. Knowles, {\it {S}pin {C}orrelations in {P}arton-{P}arton {S}cattering},
  {\em Nucl. Phys.} {\bf B310} (1988) 571.

\bibitem{Collins:1987cp}
J.~C. Collins, {\it {S}pin {C}orrelations in {M}onte {C}arlo {E}vent
  {G}enerators},  {\em Nucl. Phys.} {\bf B304} (1988) 794.

\bibitem{Richardson:2001df}
P.~Richardson, {\it {S}pin {C}orrelations in {M}onte {C}arlo {S}imulations},
  {\em JHEP} {\bf 11} (2001) 029,
  [\href{http://xxx.lanl.gov/abs/hep-ph/0110108}{{\tt hep-ph/0110108}}].

\bibitem{Gigg:2007cr}
M.~Gigg and P.~Richardson, {\it {Simulation of Beyond Standard Model Physics in
  Herwig++}},  {\em Eur. Phys. J.} {\bf C51} (2007) 989--1008,
  [\href{http://xxx.lanl.gov/abs/hep-ph/0703199}{{\tt hep-ph/0703199}}].

\bibitem{Berdine:2007uv}
D.~Berdine, N.~Kauer, and D.~Rainwater, {\it {Breakdown of the Narrow Width
  Approximation for New Physics}},  {\em Phys. Rev. Lett.} {\bf 99} (2007)
  111601, [\href{http://xxx.lanl.gov/abs/hep-ph/0703058}{{\tt
  hep-ph/0703058}}].

\bibitem{Kauer:2007zc}
N.~Kauer, {\it {Narrow-Width Approximation Limitations}},  {\em Phys. Lett.}
  {\bf B649} (2007) 413--416,
  [\href{http://xxx.lanl.gov/abs/hep-ph/0703077}{{\tt hep-ph/0703077}}].

\bibitem{Kauer:2007nt}
N.~Kauer, {\it {A Threshold-Improved Narrow-Width Approximation for BSM
  physics}},  \href{http://xxx.lanl.gov/abs/0708.1161}{{\tt 0708.1161}}.

\bibitem{Seymour:1995qg}
M.~H. Seymour, {\it {The Higgs Boson Line Shape and Perturbative Unitarity}},
  {\em Phys. Lett.} {\bf B354} (1995) 409--414,
  [\href{http://xxx.lanl.gov/abs/hep-ph/9505211}{{\tt hep-ph/9505211}}].

\bibitem{Beenakker:1996kt}
W.~Beenakker {\em et.~al.}, {\it {WW Cross-sections and Distributions}},
  \href{http://xxx.lanl.gov/abs/hep-ph/9602351}{{\tt hep-ph/9602351}}.

\bibitem{Allanach:2002nj}
B.~C. Allanach {\em et.~al.}, {\it {The Snowmass Points and Slopes: Benchmarks
  for SUSY Searches}},  \href{http://xxx.lanl.gov/abs/hep-ph/0202233}{{\tt
  hep-ph/0202233}}.

\bibitem{Porod:2003um}
W.~Porod, {\it {SPheno, A program for calculating supersymmetric spectra, SUSY
  particle decays and SUSY particle production at e+ e- colliders}},  {\em
  Comput. Phys. Commun.} {\bf 153} (2003) 275--315,
  [\href{http://xxx.lanl.gov/abs/hep-ph/0301101}{{\tt hep-ph/0301101}}].

\bibitem{Yao:2006px}
{\bf Particle Data Group} Collaboration, W.~M. Yao {\em et.~al.}, {\it {Review
  of Particle Physics}},  {\em J. Phys.} {\bf G33} (2006) 1--1232.

\bibitem{Jackson:1964zd}
J.~D. Jackson, {\it {Remarks on the phenomenological analysis of resonances}},
  {\em Nuovo Cim.} {\bf 34} (1964) 1644--1666.

\end{thebibliography}\endgroup
\end{document}